\documentclass[a4paper,two-column,12pt]{emulateapj}
\usepackage[varg]{txfonts}
\usepackage{graphicx}
\shorttitle{Lags in Seyfert 1 galaxy NGC 4593}
\shortauthors{Sriram et al.}

\begin{document}
\title{Energy Dependent time lags in the  Seyfert 1 galaxy NGC 4593}

\author{K. Sriram}
\affil{Department of Astronomy, Osmania University,
              Hyderabad 500 007, India.}

\email{astrosriram@yahoo.co.in}
\author{V. K. Agrawal\altaffilmark{1}}
          
\affil{Indian Space Research Organization, HQ., BEL Road, Bangalore 560 094, India.}

\and

\author{A. R. Rao}
\affil{Tata Institute of Fundamental Research, Mumbai 400005, India.}
         
\altaffiltext{1}{Tata Institute of Fundamental Research, Mumbai 400005, India.} 

\begin{abstract}
{We investigate the energy-time lag dependence of the source NGC 4593 using XMM-{\it Newton}/EPIC-pn data. We found that the time lag dependency is linear in nature with respect to the logarithm of different energy bands. We also investigate the frequency dependent time lags and identify that
at some frequency range (5 $\times$ 10$^{-5}$ Hz -- 2 $\times$ 10$^{-4}$ Hz) the X-ray emission is highly coherent, mildly frequency dependent and very strongly energy dependent. These observations can be explained in the frame work of the thermal Comptonization process and they indicate  a truncated accretion disk very close to the black hole.  We discuss the plausible spectral state to explain the phenomenon and conclude that the observed properties bear a close resemblance to the  intermediate state or the steep power-law state, found in galactic black hole sources.}

\end{abstract}

\section{Introduction}
It has become increasingly clear in  recent times that the central
engines of Active Galactic Nuclei (AGN) are magnified forms of Galactic
Black Hole Candidates - GBHCs \citep{mchardy06}. The scaled up X-ray
timing and spectral variability characteristics in AGNs (with respect
to GBHCs) indicate a common energy generation process in these sources.
State transition in GBHCs is a common phenomenon whereas in AGNs,
if scaled with mass, the state transition may occur in thousand to millions of
years depending on the mass of the black hole. A detailed temporal and
spectral study of a sample of AGNs, however,  may shed light on the spectral states
in AGNs.  The cross spectrum as well as power-spectral-density (PSD)
of Ark 564 \citep{arevalo06b,Mch07} indicate that the source is in  very
high state and double break in PSD and large separation between
breaks also argues against a low/hard state like that seen in  Cyg X-1
 \citep{papadakis02, done05}. The very high accretion rate inferred
in Ark564 (Romano et al. 2004) strongly suggests a Very High State
in this source.
Detailed temporal studies  indicate that the Seyfert galaxies NGC
3227, NGC 4051 and MGC-6-30-15 are in  high soft state \citep{mchardy04,
mchardy05, uttley05}. The temporal properties of NGC 3783 and NGC 4258
suggest that these sources are in low-hard state \citep{markowitz03,
markowitz05b}. An investigation of the PSD of NGC 3783 in a wider frequency
range, however, showed that the PSD is consistent with a 
soft state model (Summons et al. 2007).
Recent investigations have revealed that type 1 radio
quiet AGN and radio loud AGN spectral states are analogous to the very high
spectral state of GBHCs \citep{sobolewska09}.\\

One of the very interesting and important spectral states in GBHCs is the
very high state which is often observed during the transition from
low-hard state to high soft state \citep{mcc04, remi06, done_rev}. This state
is characterized by the presence of a steep power-law component (with power-law index
$\Gamma >2.5$) which dominates the total observed X-ray
flux, and hence this state is also referred to as the steep 
power-law (SPL) state. This state was first seen in GX 339-4 and later  observed in
many other GBHCs \citep{miya93, done_rev}.  Quasi-periodic oscillations
in the frequency range 0.1-30 Hz is also a common property of this
state \citep{mcc04}. The study of hardness-intensity diagram of
various black hole sources suggests the  onset of relativistic jet during
this transition state \citep{belloni}. A theoretical modeling of X-ray
spectra in the very high state of the XTE J1550-564 suggests that the disk
is truncated and the inner part of accretion disk is filled with hot and
compact central corona \citep{done06}.  Anti-correlated hard lags were also
discovered in XTE J1550-564 which favors a hot compact central corona in
the very high state of this source \citep{sriram07}.

It is very important to identify such states in AGNs. Phenomenologically,
one of the defining characteristics of such states is  the existence of
a confined compact corona,  which is the source of the hard X-ray power-law
 due to inverse Compton scattering of soft photons  
\citep{sunyaev80}. One of the ways to understand such a process is to study
 the time lags between two different energy bands.
If the basic process is Comptonization,  the soft photons gain energy because of fewer
 number of scattering with hot electrons present in the corona and as the energy difference increases
between two energy bands, the time lag also increases. This
energy gain time scale have been measured in a few AGNs,
NGC 7469 \citep{papadakis01}, MCG–6-30-15 \citep{vaughan03},
NGC 4051 \citep{mchardy04}, NGC 3783 \citep{markowitz05} and Mrk 110
\citep{surajit06}. Similar type of energy dependent 
lag was observed in GBHC, Cyg X-1 \citep{cui97, crary98, nowak99b}. This
clearly indicates an underlying common radiative phenomenon in both
GBHCs and AGNs. But this scenario of uniform Comptonization is, however, challenged
by the study of Fourier dependent lags. The Fourier study suggest that
the low and high frequency variation may be because of  different X-ray
emitting regions in the accretion disk and also can be explained by a
Compton cloud of radial dependence of electron density \citep{kaza97}.
Brenneman et al. (2007) have found a hard delay of the order of $\sim$230
s by cross correlating soft and hard X-ray bands in NGC 4593.
Here we try to investigate this observation further by making a detailed
energy dependent and frequency dependent analysis of the delay.

NGC 4593 is a barred spiral galaxy of Hubble type SBb, which hosts Seyfert 1 nucleus situated towards the constellation Virgo. It is a close by galaxy at a redshift of z=0.009, an apparent visual magnitude of 11.67 and harbors a super-massive
black hole of mass of the order of 8.1$\times$10$^{6}$ $M_{\odot}$ \citep{gebhardt00} - which is consistent with the poorly constrained reverberation
mapping estimate of $<$ 1.4$\times$10$^{7}$ $M_{\odot}$ (Peterson et al. 2004).
NGC 4593 is highly variable in X-ray, UV, optical and IR bands \citep{santos95} implying towards a variable continuum source close to the black hole. Multi-band spectral energy distribution of NGC 4593 suggests that the respective accretion disk is truncated and the innermost region is dominated by ADAF \citep{lu00}. The X-ray continuum is well fitted with photoabsorbed power-law above 2 keV and show spectral complexity behavior below this energy. The standard relativistic iron line is absent as well as reflection component is not seen, instead it has two narrow Fe K$\alpha$ lines at 6.4 keV and 6.97 keV in the X-ray continuum \citep{brenneman07, reynolds04}. Because of missing relativistic spectral features in X-ray band, the geometry of NGC 4593 is speculated to be different from traditional geometry of AGNs or relativistic features may be buried in the noise as discussed by Reynolds et al. (2004).

In this paper we make a detailed investigations of temporal properties of narrow line Seyfert 1 galaxy NGC 4593 using $\sim$ 70~ks data obtained by XMM-{\it Newton} observatory.

\section{Data Reduction and Analysis }
NGC 4593 was observed on 2002 June 23 by XMM-{\it Newton} satellite.
We have used EPIC pn camera data during which it was operating in
small window mode with  medium filter. We have taken ODF (Observation
Data Files) to obtain the light curves of the source. Source photons were
extracted from 40 $\times$ 40 $arcsec^{2}$ region and equal area is
taken for collecting background photons, away from the source. The {\it
epatplot} task gave no indication of pile-up during the observation. 
We have selected single and double events for the analysis (PATTERN$\la$4 and FLAG=0). The light curves are binned for 50 s and XRONOS package is used
for studying the temporal behavior of the source.\\

\subsection{Energy Dependent lag}
\begin{figure}[ht]
\includegraphics[height=9cm,width=5cm,angle=-90]{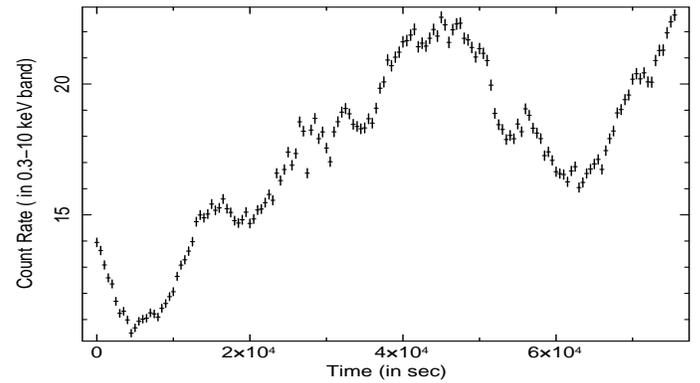}
     \caption{XMM-{\it Newton} PN 0.3-10 keV light curve of NGC 4593 binned at 50 sec.} 
       \label{Fig1}
 \end{figure}

\begin{figure}[ht]
\includegraphics[height=9cm,width=5cm,angle=-90]{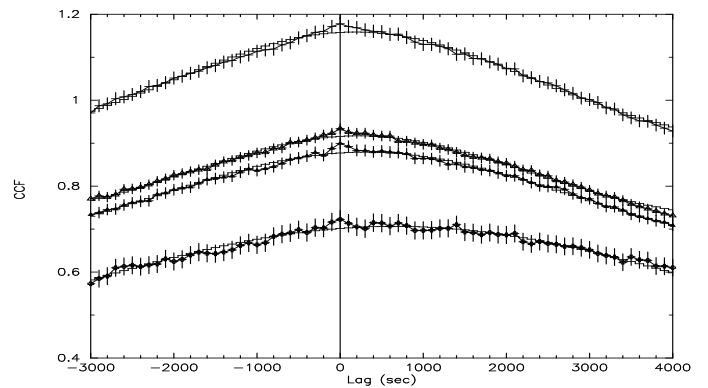}
     \caption{The cross correlation function (CCF) in various energy bands with 
respect to the soft band (0.30 -- 0.50 keV) is shown for NGC 4593. 
The energy bands are 0.50 -- 0.80 keV, 0.80 -- 1.50 keV, 1.50 -- 4.0 keV,
4.0 -- 10.0 keV, respectively from the top.
The first (top) CCF is vertically shifted up by 0.3 for clarity. The
continuous lines are Gaussian fits and the vertical line at zero
highlights the observed delay.} 
       \label{Fig2}
 \end{figure}

\begin{figure}[ht]
\centering
\includegraphics[height=9cm,width=4cm,angle=-90]{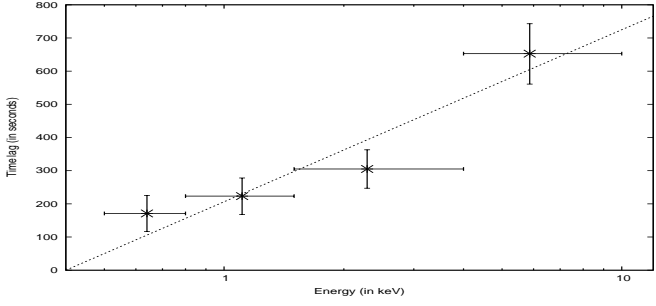}
     \caption{The observed time lags are plotted vs energy for NGC 4593.
The straight  line corresponds to thermal Comptonization model fit (see text).} 
       \label{Fig3}
 \end{figure}

We have extracted the light curves in five different energy bands (0.3-0.5
keV, 0.5-0.8 keV, 0.8-1.5 keV, 1.5-4.0 keV and 4.0-10 keV). During the
observation the source was highly variable and showed two dips separated
by a steep increase. Initially the average count rate was $\sim 12 s^{-1}$
followed by a steep increase to $\sim 25 s^{-1}$.
This is followed by a dip to around $\sim 18 s^{-1}$ and settled
down at $\sim 25 s^{-1}$ (see Fig. ~\ref{Fig1}). To know the time lag in
various energy bands, we have used {\it crosscor} program to evaluate
the CCF and the respective delay in the various energy bands. We have cross
correlated the hard X-ray bands with the soft X-ray
band  (0.3-0.5 keV) and the results are shown in Fig. ~\ref{Fig2}. 
A Gaussian model was used to evaluate the time lags and errors are estimated using the criteria $\Delta$$\chi^{2}$=4 \citep[see][]{surajit06} (see Table 1). The mean energy is given in Table 1 is the weighted mean for all the photons in each energy band. It was found that the magnitude of the lag is increasing as the
energy bands are becoming harder with respect to 0.3-0.5 keV band. It
clearly suggests that the photons take relatively more time to
become relatively more harder, most probably due to more number of
scatterings (Fig. ~\ref{Fig3}).  

{\begin{table}[ht]
\label{table}
\centering
\linespread{0.02}
\caption{Details of the selected energy bands and time lags.} 
\begin{tabular}{lll}
\hline
Correlated energy bands & mean energy (keV)& Time lag (sec) \\
\hline
0.30-0.50 vs 0.50-0.80 keV &0.64&  171.00$\pm$54.0 \\
0.30-0.50 vs 0.80-1.50 keV &1.11&  223.43$\pm$55.0 \\
0.30-0.50 vs 1.50-4.0 keV &2.29&  305.00$\pm$58.5 \\
0.30-0.50 vs 4.0-10.0 keV &5.87&  652.91$\pm$91.0 \\
\hline
\end{tabular}
\end{table}

\subsection{Frequency Dependent Lag}
If X$_1$(f) and X$_2$(f) is Fourier transform of two light curves  in
energy bands E$_1$ and E$_2$. Then cross spectrum is defined as C(f)
= X$_1^*$(f)X$_2$(f). Time lag at frequency f is given  by $\delta t =
arg[C(f)]/2 \pi f $. If S$_1$(f) and S$_2$(f) is source power in two different energy bands and N$_1$(f) and N$_1$(f) is noise spectrum in same bands then  coherence function can be defined as

\begin{equation}
\gamma ^ 2 (f)  = \frac{|\langle C(f) \rangle|^2-N^2}{\langle |S_1(f)|^2\rangle \langle|S_2(f)|^2\rangle}
\end{equation}

where
\begin{equation}
N^2 = [|S_1(f)|^2|N_2(f)|^2+|S_2(f)|^2|N_1(f)|^2+|N_1(f)|^2|N_2(f)|^2] / m
\end{equation}

where m is the number of independent measurements (Vaughan \& Nowak 1997).
The error in coherence is calculated by the method described in
 \citep{nowak99a, nowak99b}, and the error in the lags are calculated using the method 
described in Bendat \& Piersol (1986). To estimate error bars  on coherence we used high power and high measured coherence limit. We re-binned the cross spectral density logarithmically and then computed the coherence and lag. In Fig. ~\ref{Fig4}, we show coherence between 0.3-0.5 keV and 0.8-1.5 keV band light curves as function of frequency. It is clear from this figure that coherence drops above $5 \times 10^{-4}$ Hz. Power spectrum also becomes noisy above this frequency (see Fig. ~\ref{Fig5}). Hence we investigate the nature of lag spectrum in the frequency range 5 $\times$ 10$^{-5}$ Hz and 5 $\times$ 10$^{-4}$ Hz. We also compute cross-spectra between two different energy bands 0.3-0.5 keV vs 0.8-1.5 a keV and 0.3-0.5 keV vs 4.0 keV - 10.0 keV. Frequency dependent lag is shown in Fig. ~\ref{Fig6} and it is evident in the figure that lag scales up with energy separation.

\begin{figure}[ht]
\includegraphics[height=8cm,width=4cm,angle=-90]{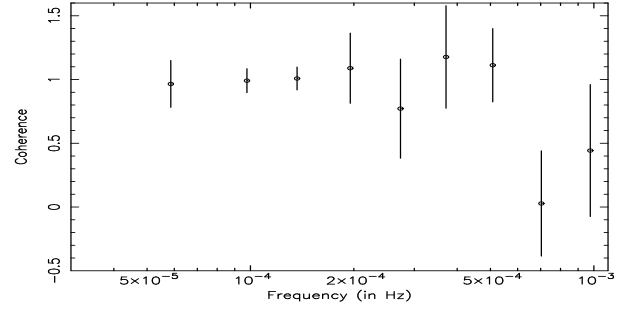}
     \caption{Coherence function calculated using energy bands 0.3-0.5 keV and 0.8-1.5 keV.} 
\label{Fig4}
\end{figure}
\begin{figure}[ht]
\includegraphics[height=8cm,width=4cm,angle=-90]{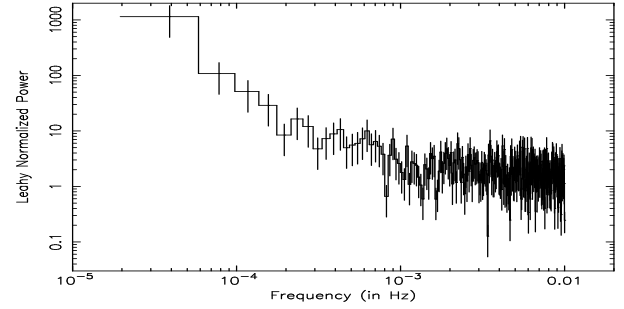}\\
     \caption{The  Power Density Spectrum (PDS) of NGC 4593 in 0.8-1.5 keV energy band.} 
       \label{Fig5}
 \end{figure}
\begin{figure}[ht]
\includegraphics[height=8cm,width=4cm,angle=-90]{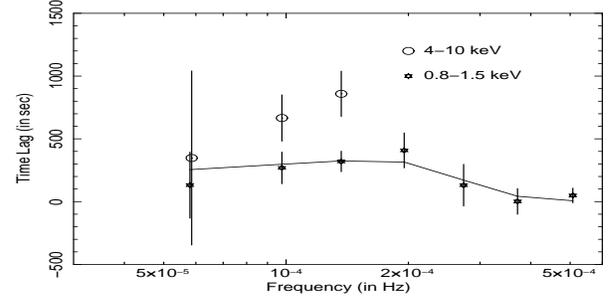}
     \caption{Frequency dependent lag between 0.3-0.5 keV vs 0.8-1.5 keV and 0.3-0.5 keV vs 4.0 keV-10.0 keV energy bands. The line shows the broken power-law fit (see text).}
 \label{Fig6}
\end{figure}

\section{Discussion and Conclusion}
\subsection{Energy dependent lag from Comptonization}
In this paper we investigated the dependence of time lag on energy of observed
X-ray radiation in NGC 4593. Our results suggests that the time lag increases
with the mean energy of observed photons. A similar dependence of lag on
energy has been seen in other AGNs as well and the observed lag is
interpreted as Compton scattering time scale \citep{surajit06}. 
If this lag is due to the process of Comptonization, we can get some 
useful constraints on the geometry of the inner accretion disk.
We note here that the Comptonization model applies to the 
energy dependent lags measured at high temporal frequencies. 

 Let R be the size of the region
in which the Comptonization process is undergoing. If  $\lambda$ 
is the mean free path of the photon and $\tau_{T}$ is optical depth of the plasma
in this region, then the 
Comptonization time scale will be given by \\
\begin{equation}
t_{Comp}=\frac{\lambda}{c*max(1, \tau_{T})}
\end{equation}

 Let us assume $E_{o}$ to be the initial energy of the photon
and after N scattering it attains an energy $E_{N}$ (where N=1,2,3,4),
i.e. $E_{N}$=E$_{0}$*A$^N$, where A=1+4$\Theta$+16$\Theta^{2}$ (know as
amplification factor and $\Theta$=$kT_{e}/m_{e}c^{2}$). Then the 
Comptonization time scale can also be written in terms of time lag and N as
\begin{equation}
t_{Comp}=\frac{\tau_{lag}}{N}
\end{equation}
where $\tau_{lag}$ is the respective time lag.\\ 

The size of the region
in which the Comptonization process is undergoing is equivalent to
the distance traveled by the photons (i.e $\lambda$, valid only if
the medium is optically thin, i.e. max(1, $\tau_{T}$)$\sim$1). Using
the above two equations we get a Comptonizing region size of the
order of $\sim$6$\times$ $10^{12}$ cm (using $\tau_{lag}$=600
s, for the 0.3 -- 0.5 keV and 4 -- 10 keV bands). 

%
The observed time lags are nothing but product of Comptonization time
scale and difference of successive number of scattering, which can be given by
%
\begin{equation}
\tau_{lag}=\frac{R}{c*max(1, \tau_{T})} \frac{ln({\frac{E_{N}}E_{1}})}{ln A}
\end{equation}

The observed time lag increases
linearly with logarithmic of energy (see Fig. \ref{Fig3}). Hence the above
equation  suggests
that the Comptonization process is responsible for the observed lag.  We have
fitted the above equation to the observed energy dependent time lag
(see Fig. ~\ref{Fig3}) and found the size of hard X-ray emitting region
is $\sim$$10^{13}$ cm and when the mass is taken 7$\times$ $10^{6}$ $M_{\odot}$ black hole (Nelson et al. 2004),
R comes out to be $\sim$5 R$_{g}$.

\subsection{Disk Geometry and Time Lag spectrum}
The geometry of the accretion disk in NGC 4593 is quite peculiar on the basis
 of Fe line \citep{brenneman07, reynolds04}.  Also iron  line does not
 show significant variation on observed time span and are very narrow
 which are suggestive of absence of optically thick cold matter in
 the central 1000 R$_g$ region. The inner 1000 R$_g$ region may contain 
radiatively inefficient and hot optically thin plasma. The observed time
 lag suggests that there is also a compact central corona with size of
 $\sim$ 5 R$_g$. The derived Compton cloud size from time lag suggests that most
 of the hard X-ray photons are emitted within $\sim$3$R_{g}$ region.
 It has also been argued that NGC 4593 has a truncated accretion disk
 with a two component flow of a geometrically thin disk and an advection
dominated flow within the truncation radius, a model commonly
invoked to explain the low-hard and intermediate states of GBHs \citep{lu00}.
 The intermediate state state is characterized by a steep power law component
 and a truncated accretion disk (with the truncation radius less
 than that found in the low hard state).  Though the power-law index 
is not steeper than that found in other Seyfert galaxies \citep{brenneman07},
the available data strongly suggest a thin disk converting into
an advection dominated central part, a configuration
 similar to the low-hard and intermediate states of GBHs.

Cross spectral analysis reveals that lag observed in this source depends
upon Fourier frequency. The lag spectrum computed using energy bands
0.3-0.5 keV and 0.8-1.5 keV reveals a broad peak in the frequency range
5 $\times$ 10$^{-5}$ Hz to 5 $\times$ 10$^{-4}$ Hz (see Fig. \ref{Fig6}).  In this frequency
range, the X-ray emission at different energies are highly
coherent (see Fig. \ref{Fig4}). The lag spectrum, however,  shows
 quite strong energy dependency. The structure in
the lag spectrum can be fitted with a  broken power law with first slope
$\alpha$ = 0.3 and second slope $\beta$ = -5 with a $\chi^{2}$ = 1.5 for 3 degree of freedom (d.o.f), an improvement in $\chi^{2}$ of $\sim$10 from
a fit to a constant value. The break is observed
at frequency $f_b$ = 2.2 $\times$ 10$^{-4}$ Hz. The observed lag at
different frequencies increases with energy separation.  The lag spectrum
of Ark 564 shows step-like features (McHardy et al. 2007)
 and at higher frequencies
it can also  be fitted with a broken power-law
\citep{arevalo06b}. 
In Ark 564 also the lag
at different frequencies scales up with energy separation. 
The lags below the break frequency is a constant in NGC 4593, which
is quite different from the trend of lag decresing with increasing Fourier
frequency seen in other sources like MCG-6-30-25 (Vaughan, Fabian \& Nandra 2003),
Mkn766 (Markowitz et al. 2007), NGC 4051 (McHardy et al. 2004).
The fractional lag in this frequency range is 1 -- 6\%, which is
similar to the fractional lag seen in the very high state of Cyg X-1.\\

If the energy dependent lag is mainly due to Comptonization, then the
bulk of the variability (confined to below 2  $\times$ 10$^{-4}$ Hz, see Fig. \ref{Fig6}) is nearly frequency independent. It is quite
possible that there is another variability component operating particularly
above the break frequency $f_b$ = 2.2 $\times$ 10$^{-4}$ Hz, possibly 
from accretion disk fluctuations propagating from outside (see below).
The spectral analysis of NGC 4593 suggests
that the soft excess too can be modelled as due to Comptonization from
a cooler plasma (Brenneman et al. 2007). It is quite possible that there is a hot compact central corona within a few Swartzchild radius
(responsible for the bulk of the radiation and giving rise to the frrequency 
independent delays), surrounded by an optically thin
plasma cloud extending to several hundred Swartzchild radius
 (responsible for the soft excess). There is a possibility that electron density in optically
thin corona is varying with radius. It was suggested that if corona
has non-uniform electron density then observed frequency dependence lag
can be explained \citep{kaza97}. There is also a possibility that there
exists several different patches of hot Compton cloud \citep{sobo04}
giving rise different time lags at different time scales. \\

Another promising model which explains the observed lag-spectrum in Mrk 335 
\citep{arevalo08} is fluctuations propagation model \citep{lyu97, kotov01},
which can also give energy dependent delays. This model assumes that 
fluctuations in the accretion flow are produced at different radial
 distances from putative black hole. These fluctuations propagates on
 viscous time scales and also assumes that spectra hardens toward the center.
 Kotov et al. (2001) has computed phase lags for different scenarios 
which can explain the time lags caused by accretion. The lag spectrum
 observed in NGC 4593 is similar to   that obtained for freely falling
 blob model(see Fig. 9 of Kotov et al. 2001), but the strong
energy dependency, however, favors a Comptonization model.

\acknowledgements
  We thank the anonymous referee for the very useful comments. This work has made use of observations obtained with XMM-{\it Newton}, an ESA science mission with instrumentation and contributions directly funded by ESA member states and the US (NASA). The authors are grateful to Surajit Dasgupta for useful suggestions. K.S is supported by UGC through RFSMS scheme and thankful to TIFR for providing the facility to carry out the work.

\end{document}